\documentclass{article}
\usepackage{frascatiphys,here,graphicx,subfigure}
\usepackage{amssymb}
\def\mpipi{M_{\pi\pi}}
\begin{document}
\title{ 
THE GLUEBALL AMONG THE LIGHT SCALAR MESONS
}
\author{
Wolfgang Ochs        \\
{\em Max-Planck-Institutf\"ur Physik, F\"ohringer Ring 6, D-80805 M\"unchen,
Germany} \\
}
\maketitle
\baselineskip=11.6pt
\begin{abstract}
The lightest gluonic meson is expected with $J^{PC}=0^{++}$, calculations in 
full QCD point towards a mass of around 1 GeV. 
The interpretation of the scalar meson spectrum is hindered
as some states are rather broad.
In a largely model-independent analysis 
of $\pi^+\pi^-\to \pi^+\pi^-,~\pi^0\pi^0$ scattering in the
region 600-1800 MeV a unique solution for the isoscalar S-wave is obtained.
The resonances $f_0(980),\ f_0(1500)$ and the broad 
$f_0(600)$ or ``$\sigma$'' are clearly identified whereas $f_0(1370)$ is not
seen at the level  $B(f_0(1370)\to \pi\pi)\gtrsim 10\%$. 
Arguments for the broad state
to be a glueball are recalled. We see no contradiction with the reported 
large $B(\sigma\to \gamma\gamma)$ and propose 
some further experimental tests.
 \end{abstract}
\baselineskip=14pt
\section{QCD predictions for the lightest glueball}
The existence of gluonic mesons belongs to the early predictions of QCD
and first scenarios have been developed back in 1975\cite{fm}.
Today, quantitative results are available from \\
1. Lattice QCD: 
In full QCD both glue and $q\bar q$ states couple to the flavour singlet
$0^{++}$ states and first ``unquenched'' results for the lightest gluonic 
state point
towards a mass of around 1 GeV\cite{michael}. 
This is a considerably lower mass value than what is
obtained in the pure Yang Mills theory for gluons (quenched approximation)
where the lightest glueball is found at masses around 1700 MeV (recent
review\cite{neile}). Further studies concerning the dependence on lattice spacing
and the quark mass appear important.\\
2. QCD sum rules: Results on the scalar glueball and various decays are
obtained in\cite{veneziano}. The lightest gluonic state is found in
the mass range (750-1000) MeV with a decay width of (300-1000) MeV
into $\pi\pi$ and the width into $\gamma\gamma$ of (0.2-0.3) keV.
Other analyses find similar or slightly higher masses $(1250\pm200)$ MeV
for the lightest glueball\cite{steele}.
\section{The scalar meson spectrum and its interpretation}
In the search for glueballs one attempts to group the scalar mesons
into flavour multiplets (either $q\bar q$
or tetraquarks) and to identify supernumerous states.
The existence of such states could be a hint for glueballs either pure or
mixed with $q\bar q$ isoscalars.
In other experimental activities one looks for states which are
enhanced in ``gluon rich'' processes and are suppressed in $\gamma\gamma$
processes.

The lightest isoscalar states listed in the particle data group\cite{pdg}
are
\begin{equation}
f_0(600) ({\rm or}~
\sigma),~f_0(980),~f_0(1370) (?),~f_0(1500),~f_0(1710),~f_0(2080),
\label{scalars} 
\end{equation}
where the question mark behind $f_0(1370)$ will be explained below.
There are different routes to group these states into multiplets
together with $a_0$ and $K^*_0$ states.

In a popular approach the two lightest isoscalars in (\ref{scalars})
are combined with $\kappa(800)$ and $a_0(980)$ to form the lightest nonet,
either of $q\bar q$ or of $qq-\bar q\bar q$ type. Then the next higher
multiplet from $q\bar q$ would include $a_0(1450),~K^*_0(1430)$; near these
masses three isoscalars are found in the list (\ref{scalars}) at
1370, 1500 and 1710 MeV and this suggests to consider these three
mesons as mixtures of the two members of the $q\bar q$ nonet and one glueball
(for an early reference, see\cite{ac}). 

A potential problem in this scheme for the glueball is the very existence of
$f_0(1370)$, otherwise there is no supernumerous state in this mass range.
Some problems with this state will be discussed below, see also the
review\cite{klemptrev}. 
The low mass multiplet depends on the existence
of $\kappa$ which we consider as not beyond any doubt:
its observed phase motion is rather weak and it is markedly
different from the one of $``\sigma"$, see below.

There are other approaches for the classification of the scalar mesons
where $f_0(980)$ is the lightest $q\bar q$ scalar.
In the scheme we prefer\cite{mo}
the lightest $q\bar q$ nonet contains $f_0(980),~f_0(1500)$ together with 
$a_0(1450),~K^*_0(1430)$. The supernumerous state $f_0(600)$, called
previously $f_0(400-1200)$, corresponds to a very broad
object which extends from $\pi\pi$ threshold up to about 2 GeV
and is interpreted as largely gluonic. No separate
$f_0(1370)$ is introduced, nor $\kappa(800)$. 
Our classification is consistent with various findings on production and 
decay processes including $D,D_s,B$ and $J/\psi$
decays\cite{mo,momontp,mobdecay}. 

Related
schemes are the Bonn model\cite{klempt} with a similar mixing scheme for
the isoscalars and the K-matrix model\cite{anis} which finds a similar
classification (but with $f_0(1370)$ included) and a broad glueball, 
centered at the higher masses around 1500 MeV. 
\section{Study of $\pi\pi$ scattering from 600 to 1800 MeV}

\subsection{Selection of the physical solution for $m_{\pi\pi}>1000$ MeV}

We are interested here in particular in the problem of $f_0(1370)$ and also 
in the behaviour of the broad ``background'' which is related to $f_0(600)$
or ``$\sigma$'', alias $f_0(400-1200)$ and describe the results from an ongoing analysis (see also
\cite{womont}). 

Information on $\pi\pi$ scattering can be obtained from production
experiments like $\pi p \to \pi \pi n$ by isolating the contribution of the
one-pion-exchange process. In an unpolarised target experiment
these amplitudes can be extracted by using dynamical assumptions, such as
``spin and phase coherence'', which have been tested by experiments with
polarised target. At the level of the process $\pi\pi\to\pi\pi$ in different
charge states one measures the distribution in scattering angle,
$z=\cos\theta^*$, or their moments $\langle Y^L_M \rangle$, in a sequence of
mass intervals. The $\pi\pi$ 
partial wave amplitudes $S,P,D,F,\ldots$ can be obtained
in each bin from the measured moments up to the overall phase and a discrete
ambiguity (characterised by the ``Barrelet Zeros''). The overall phase can
be fixed by fitting a Breit Wigner amplitude for the leading resonances
$\rho,~f_2(1270)$ and $\rho_3(1690)$ to the experimental 
moments $\langle Y^2_0 \rangle$,
 $\langle Y^4_0 \rangle$ and  $\langle Y^6_0 \rangle$ respectively.

Phase shift analyses of this type for $\pi^+\pi^-$ scattering have been performed
by the CERN-Munich group: an analysis guided by a global resonance fit (CM-I
\cite{cm}) and a fully energy-independent analysis by CM-II\cite{cm2} and by
Estabrooks and Martin\cite{em}; the latter two 
analyses found 4 different solutions above 1 GeV in mass. Up to 1400 MeV a
unique solution has been found\cite{kpy} using results from polarised
target 
and unitarity. Two solutions remain above 1400 MeV,
classified according to Barrelet zeros in\cite{cm2} as ($---$) and ($-+-$).
corresponding to sols. A,C in\cite{em}.

A new result has been added recently\cite{womont} by 
the construction of the isoscalar S wave $S_0$ from the  $\pi^+\pi^-\to
\pi^0\pi^0$ data (GAMS collaboration\cite{gams}) 
and the $I=2$ scattering data. 
This $S_0$ wave shows a
qualitatively similar behaviour to $S_0$ obtained from 
$\pi^+\pi^-\to \pi^+\pi^-$ scattering above, namely a resonance circle
in the complex plane (Argand diagram) related to $f_0(1500)$ above a slowly
moving circular background amplitude. This has lead us to select the solution
($-+-$) as unique solution. We relate the differences in the two results 
to systematic errors introduced through the
overall phase and the $S_2$ wave, but these are only slowly varying
effects as function of mass.

\begin{figure*}[t]
\begin{center}
\begin{tabular}{@{}lll}
\includegraphics*[angle=-90,width=5cm,bbllx=3cm,bblly=1.5cm,bburx=19.5cm,%
bbury=19.5cm]{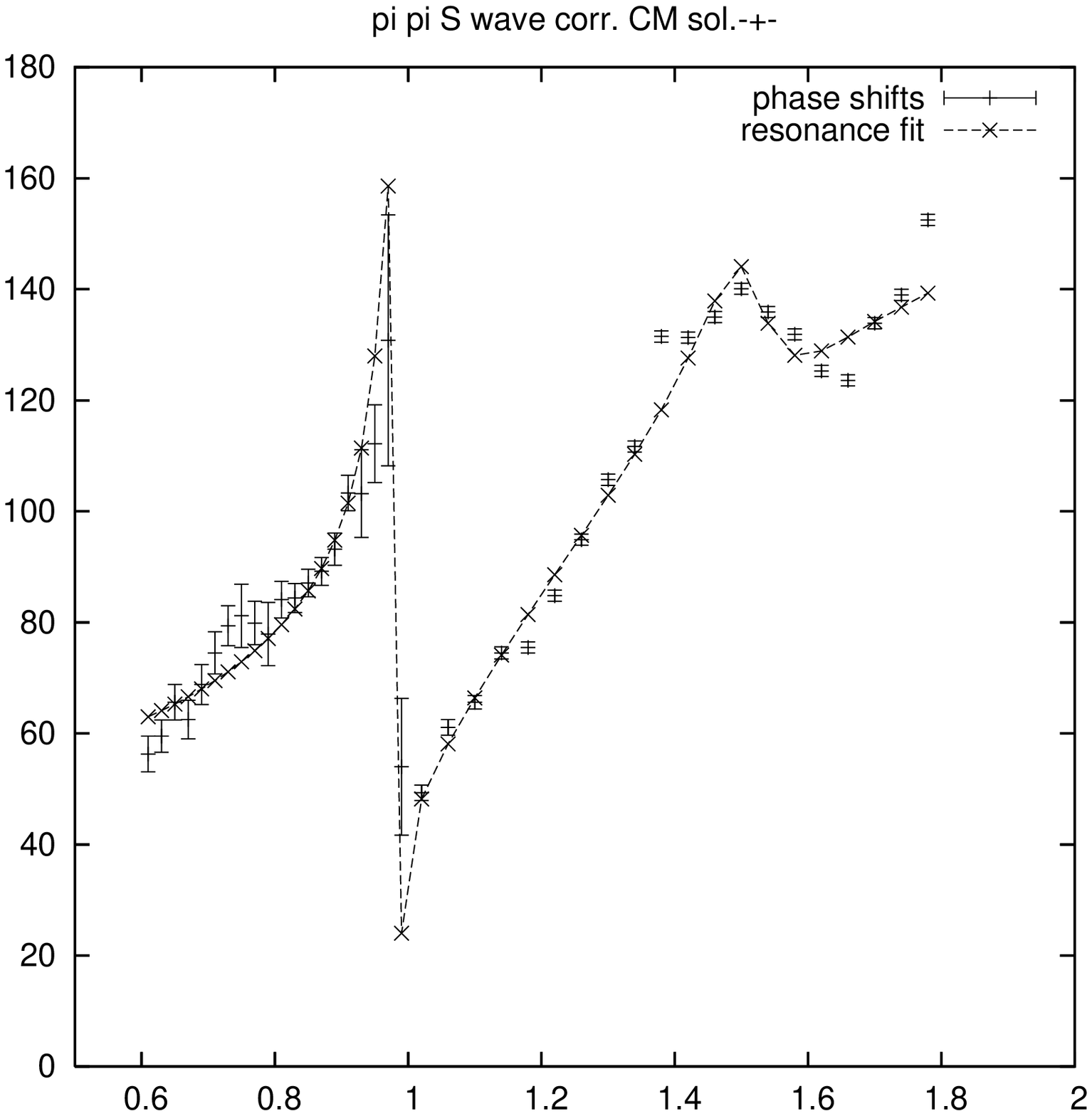},
\includegraphics*[angle=-90,width=5cm,bbllx=3cm,bblly=1.5cm,bburx=19.5cm,%
bbury=19.5cm]{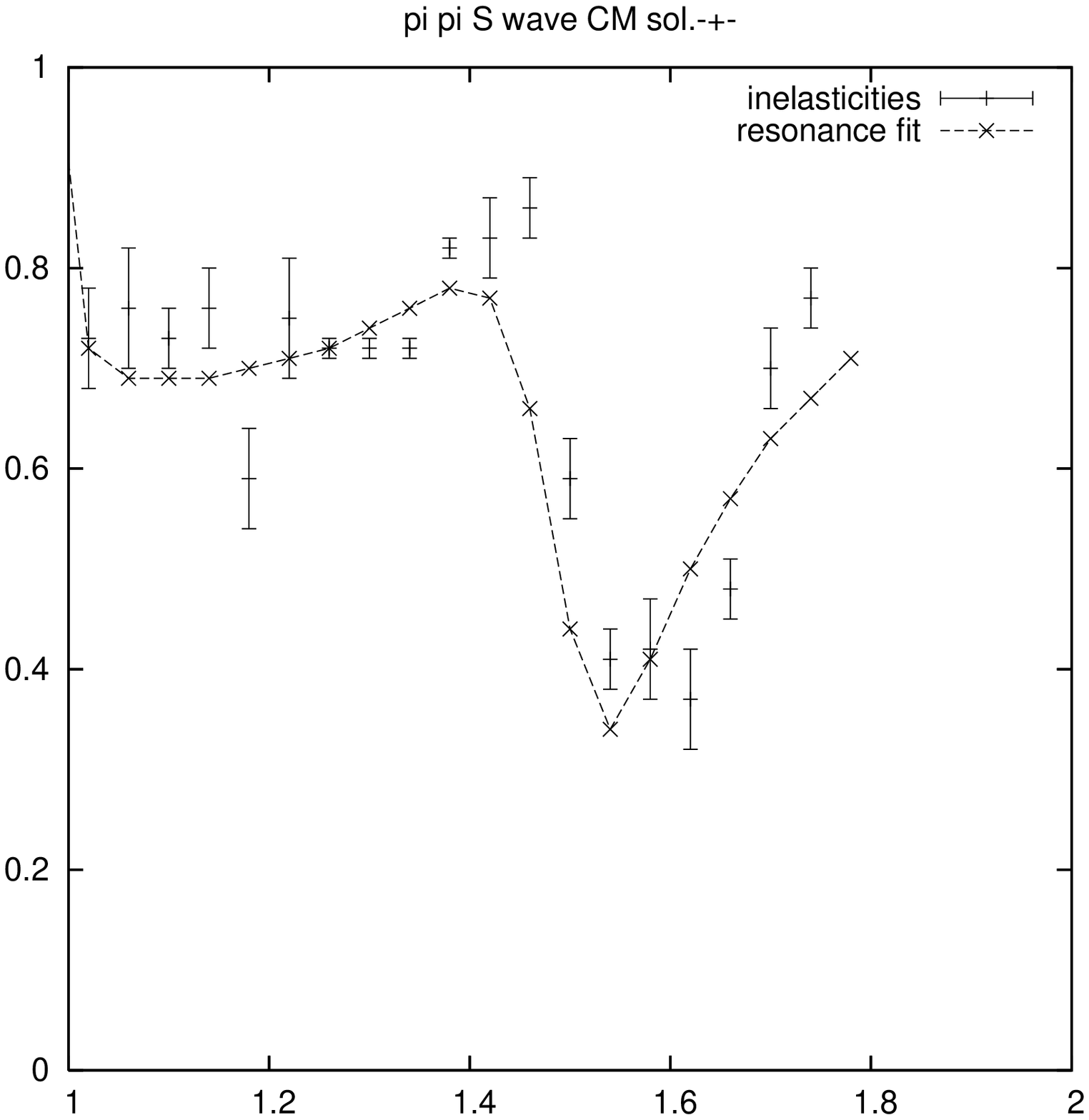} \\
\includegraphics*[angle=-90,width=5cm,bbllx=3cm,bblly=1.5cm,bburx=19.5cm,
bbury=19.5cm]{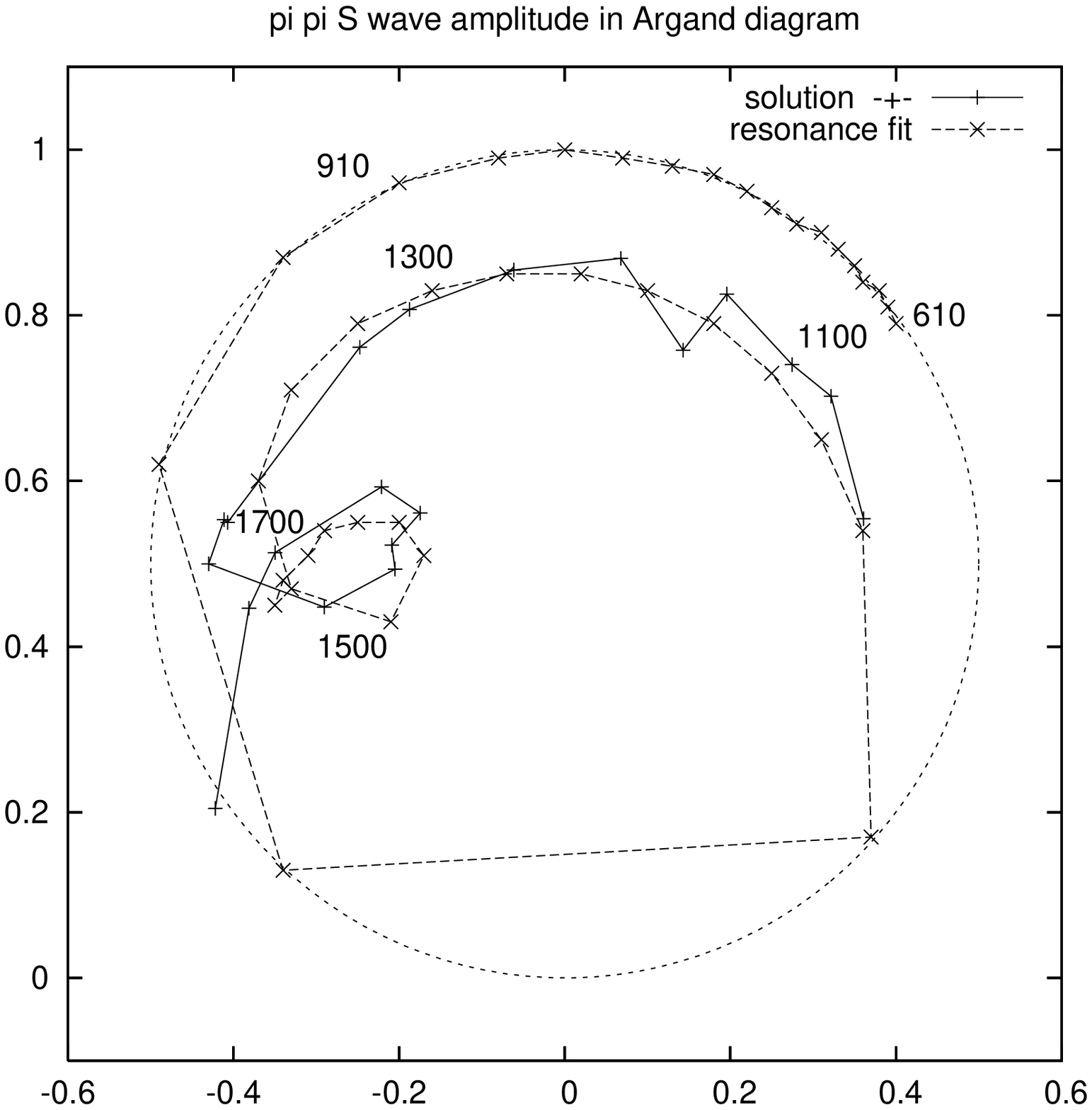},
\includegraphics*[angle=-90,width=5cm,bbllx=3cm,bblly=1.5cm,bburx=19.5cm,
bbury=19.5cm]{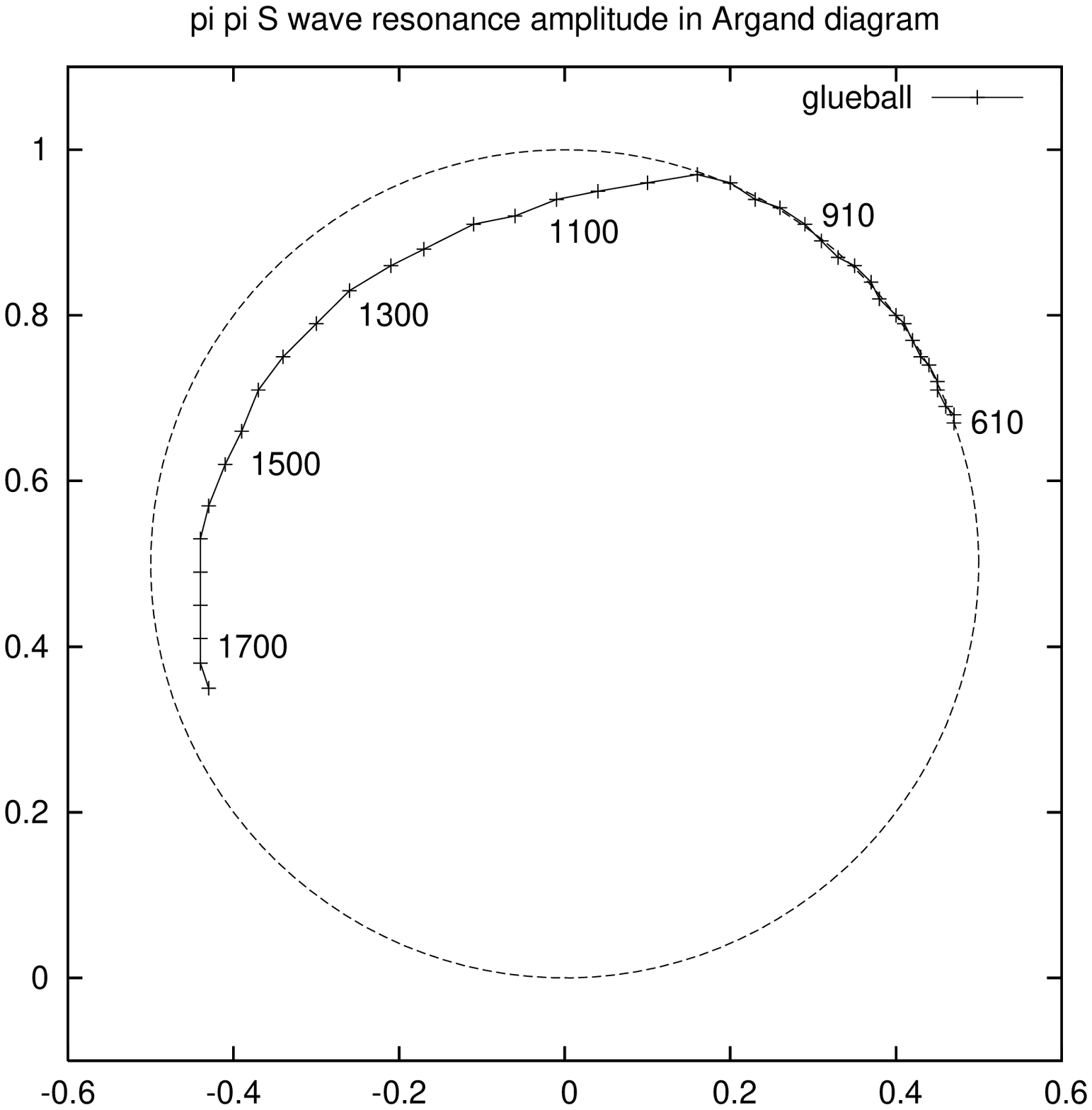}\\
\end{tabular}
\end{center}
\caption{Resonance fit Eq. (2) in comparison with data on the corrected 
$S_0$ wave (CM-I/II): 
phase shifts $\delta_0^0$ and inelasticity $\eta^0_0$;
Argand diagram (Im $S_0$ vs. Re $S_0$); 
on rhs: broad component
$f_0(600)/\sigma$ from the fit.}
\label{fig:resonances}
\end{figure*}

\subsection{Resonance fit to the isoscalar S wave}
The resulting amplitude $S_0(-+-)=(\eta^0_0\exp(2i\delta^0_0)-1)/2i$ 
is shown in Fig. \ref{fig:resonances} using the CM-II data after
correction for the more recent $I=2$ amplitudes. 
The curves refer to a fit of the data (CM-II for $M_{\pi\pi}>1$ GeV, 
CM-I for $M_{\pi\pi}<1$ GeV)
to an S-matrix in the space of 3 reaction channels
($\pi\pi,K\bar K, 4\pi$) as product of individual S-matrices for resonances
$S_R=1+2i T_R$
\begin{eqnarray}
S&=&S_{f_0(980)}S_{f_0(1500)}S_{\rm broad}\\
T_R&=& [M_0^2-\mpipi^2-i(\rho_1 g_1^2 +\rho_2 g_2^2+ \rho_3 g_3^2)]^{-1}
   \nonumber\\
   & &   \phantom{xxx}\times\rho^{\frac{1}{2}T} (g_ig_j)\rho^\frac{1}{2}
    \label{smatrix}
\end{eqnarray}
where $\rho_i=2k_i/\sqrt{s}$. As can be seen in Fig. \ref{fig:resonances} 
the fit including 3 resonances gives a reasonable description of the data.  
For $f_0(1500)$ the fit parameters $M_0= 1510~{\rm MeV},~\Gamma_{tot}=
88~{\rm MeV},~B(f_0\to\pi\pi)=38\%$ are obtained in remarkable agreement
to the PDG numbers, despite the different approaches involved. 

\subsection{Note on $f_0(600),~\kappa(800)$ and $f_0(1370)$}
The broad object is also described
by a resonance form with mass parameter $M_0\sim1100$ MeV and width
$\Gamma\sim1450$ MeV. The elastic width is about  85\% whereas the GAMS data
suggest rather a smaller value around 70\%. More details will be given
elsewhere. This parametrisation is also shown at the lower rhs of Fig.
\ref{fig:resonances}. It describes about 3/4 of the full resonance circle.
The Breit Wigner mass parameter $M_0$ 
denotes the mass where the amplitude is purely
imaginary. It is different from the pole mass which is referred to as
resonance mass. This mass value appears to be considerably lower 
and requires a
more careful study of the line shape in the denominator of (\ref{smatrix}).

In any case, the data in Fig. \ref{fig:resonances} suggest there is 
evidence for a broad state in $\pi\pi$,
centered around 1000 MeV along the physical region 
and what is called $f_0(600)$ or $\sigma$ refers to the same state,
there cannot be two states.

We also note here that the $\pi\pi$ scattering looks considerably different
from elastic $K\pi$ scattering in that the phase of the ``background'' 
found in the analysis of the LASS data\cite{lass} moves more slowly
staying below 90$^\circ$ always. The existence of $\kappa$ 
would become evident if the phase passed through $90^\circ$ in forming a
circle as in case of $\sigma$.

We note that the data presented in Fig. \ref{fig:resonances} 
do not give any indication of 
the existence of $f_0(1370)$ which would show up as
a second circle in the Argand diagram with respective signals 
in $\eta^0_0$ and $\delta_0^0$. In fact, 
none of the energy-independent bin by bin analyses of the CM
or CKM data\cite{cm,em,cm2,klr} nor of the GAMS data\cite{gams,womont} 
gave such an indication. From our analysis 
we exclude an additional state  with branching ratio
$B(f_0(1370)\to \pi\pi)\gtrsim 0.1$ near 1370 MeV (this would correspond to
a circle of  diameter 0.1). 

These results from the bin-by-bin analysis are in apparent conflict 
with two other analyses presented at this
conference\cite{bugg,sarantsev}. In both
studies 
CM-I moments as well as various other data sets from 3 body final states, 
have been fitted by model
amplitudes with resonances in all relevant partial waves.
The amplitude $S_0$
by Bugg\cite{bugg} shows $f_0(1370)$ as an extra circle 
of diameter 0.25 whereas Sarantsev's Argand diagram\cite{sarantsev} 
shows no extra circle but an effect in the phase movement. Obviously, these
discrepancies need to be understood.
 
\section{Glueball interpretation of the broad object $f_0(600)$}
The following arguments are in favour of this state to be a
glueball\cite{mo,momontp,mobdecay}.\\
1. This state is produced in almost all ``gluon rich'' processes, including
central production $pp\to p(\pi\pi)p$, $p\bar p\to 3\pi$, $J/\psi\to
\gamma \pi\pi(?)$, $\gamma K\bar K,\gamma 4\pi$, $\psi'\to \psi\pi\pi$,
$\Upsilon'',\Upsilon'\to \Upsilon \pi\pi$ and finally $B\to K\pi\pi,
B\to K\bar K K$ related to $b\to sg$. The high mass tail above 1 GeV is seen
as ``background'' in $J/\psi \to \gamma K\bar K$ and in $B$ decay
channels where it leads to striking interference phenomena with
$f_0(1500)$\cite{mobdecay}.\\
2. Within our classification scheme\cite{mo} without $\kappa$ 
and $f_0(1370)$ the state
$f_0(600)$ is supernumerous.\\
3. The mass and large width is in agreement with the QCD sum rule results
and also with the first results from unquenched lattice QCD.\\
4. Suppression in $\gamma\gamma$ production.\\
Recently, the radiative width $\Gamma(f_0(600)\to\gamma\gamma)=(4.1\pm 0.3)$
keV has been determined by Pennington\cite{pennington} from the process
$\gamma\gamma\to \pi\pi$. As this
number is larger than expected for glueballs (see Sect.1), he concluded 
this state ``unlikely to be gluonic''. Similar results on this width 
are obtained 
by\cite{mmno,oller}. A resolution of this conflict has been suggested in a
recent paper\cite{mmno}.

It is argued that the phenomenology of $\gamma\gamma\to \pi\pi$ 
at low energies is different from the one at high energies. At low energies,
few 100 MeV above threshold, the photons couple to the charged pions 
and the Born term with one pion exchange dominates
in $\gamma\gamma\to \pi^+\pi^-$, in addition there is a contribution from 
$\pi^+\pi^-$ rescattering. Explicit models with $\pi\pi$ scattering as input
and with $f_0(600)$ pole, can explain the low energy
processes\cite{mennessier,goble}, also calculations in $\chi PT$ with
non-resonant $\pi\pi$ scattering at low energies
\cite{chipt}. In this case of the rescattering contribution, a resonance
decaying into $\pi\pi$ would also decay into $\gamma\gamma$ irrespective of
the constituent nature of the state.

At high energies, the photons do resolve the constituents of the produced
resonances: for example, the radiative widths of tensor mesons $f_2,f_2',a_2$
in the region 1200-1500 MeV follow the expectations
from a $q\bar q$ state.
   
In the model by Mennessier\cite{mennessier} the low energy rescattering and
the high energy ``direct'' component relating to the constituents 
are added; the unitarization keeps the
validity of Watson's theorem. A fit of the data 
at the lower energies $\mpipi<550$ MeV provides an estimate of the direct
contribution from its deviation from the rescattering term. This yields 
$\Gamma(f_0(600)\to\gamma\gamma)|_{direct}\approx 0.3$ keV ($\pm 50\%$),
alternatively, one can express this result as upper limit 
$\Gamma(f_0(600)\to\gamma\gamma)|_{direct}<0.5$ keV (90\%CL).
This result implies that there is no contradiction with a gluonic
interpretation of $f_0(600)$.

Finally, we express some expectations for experiment which follow from this
interpretation.\\
1. Because of its large width the state $f_0(600)$ overlaps with both
physical regions. Whereas the low energy region is governed by hadronic
rescattering there is the transition to high energies with a resolution of
the constituents. Therefore we expect that for increasing mass
$\mpipi\gtrsim 1$ GeV the decay fraction
$f_0(600)\to \gamma\gamma$ decreases strongly relative to $f_0(600)\to
\pi\pi$ in consequence of the weak intrinsic coupling of the glueball to
$\gamma\gamma$ by an order of magnitude.   \\
2. In processes with virtual photons the $\pi\pi$ rescattering
contribution should be suppressed with respect to the direct $q\bar q$ coupling 
contribution because
of the pion formfactor. This could result in a relative suppression of
$f_0(600)$ production at low $\pi\pi$ mass 
with respect to $f_0(980)$ if the latter state is
dominantly $q\bar q$; this should hold for both space like
($\gamma_V\gamma\to \pi\pi$) and time like photons ($\gamma_V\to \pi \pi
\gamma$).  

In this way the study of the $\pi\pi~S$ wave cross section 
in two-photon processes (or its upper
limit obtained using the positivity of the 
density matrix\cite{positivity}) could provide
new clues on the interpretation of the broad state $f_0(600)$.

\section{Acknowledgements}
I would like to thank Gerard Mennessier, Stephan Narison and Peter 
Minkowski for discussions and the collaboration on subjects of this talk.

\end{document}